\documentclass[12pt]{article}
\makeatletter

\def\ltap{\raisebox{-.4ex}{\rlap{$\sim$}} \raisebox{.4ex}{$<$}}

\newcommand{\drawsquare}[2]{\hbox{%
\rule{#2pt}{#1pt}\hskip-#2pt
\rule{#1pt}{#2pt}\hskip-#1pt
\rule[#1pt]{#1pt}{#2pt}}\rule[#1pt]{#2pt}{#2pt}\hskip-#2pt
\rule{#2pt}{#1pt}}

\def\ltap{\raisebox{-.4ex}{\rlap{$\sim$}} \raisebox{.4ex}{$<$}}

\newcommand{\Yfund}{\raisebox{-.5pt}{\drawsquare{6.5}{0.4}}}

\newfont{\sss}{cmss17 scaled 1800}

\begin{document}
\begin{titlepage}
\today          \hfill
\begin{center}
\hfill    LBNL-41856 \\
\hfill    UCB-PTH-97/65 \\
{\large \bf 
Getting the Supersymmetric Unification Scale from 
Quantum Confinement with Chiral Symmetry Breaking}
\footnote{This work was supported in part
by the Director, Office of Energy
Research, Office of High Energy and Nuclear Physics,
Division of High
Energy Physics of the U.S. Department of Energy
under Contract
DE-AC03-76SF00098.
The author was also supported by NSERC.}

\vskip .25in

M. Graesser
\footnote{email: graesser@thsrv.lbl.gov}

{\em Theoretical Physics Group\\
    Lawrence Berkeley National Laboratory\\
    and \\
    Department of Physics \\
      University of California\\
    Berkeley, California 94720}
\end{center}

\begin{abstract}
Two models which generate the 
supersymmetric Grand Unification Scale from the 
strong dynamics of an  
additional gauge group are presented. 
The particle content is chosen such that this group 
confines with chiral symmetry breaking.
Fields that are usually introduced to break the Grand Unified 
group appear instead as composite degrees of freedom and can acquire 
vacuum expectation values due to the confining dynamics. 
The models     
implement known solutions to the doublet-triplet
splitting problem. The $SO(10)$ model only requires one higher dimensional
representation, an adjoint.
The dangerous coloured Higgsino-mediated proton decay operator is 
naturally suppressed in this model to a phenomenologically 
interesting level. Neither model requires the presence of gauge singlets.
Both models are only technically natural.
\end{abstract}
\end{titlepage}
\renewcommand{\thepage}{\roman{page}}
\setcounter{page}{2}
\mbox{ }

\vskip 1in

\begin{center}
{\bf Disclaimer}
\end{center}

\vskip .2in

\begin{scriptsize}
\begin{quotation}
This document was prepared as an account of work sponsored by the
United
States Government. While this document is believed to contain
correct
 information, neither the United States Government nor any agency
thereof, nor The Regents of the University of California, nor any
of their
employees, makes any warranty, express or implied, or assumes any legal
liability or responsibility for the accuracy, completeness,
or usefulness
of any information, apparatus, product, or process disclosed, or
represents that its use would not infringe privately owned rights.
Reference herein
to any specific commercial products process, or service by
its trade name,
trademark, manufacturer, or otherwise, does not necessarily
constitute or
imply its endorsement, recommendation, or favoring by the
United States
Government or any agency thereof, or The Regents of the
University of
California.  The views and opinions of authors expressed herein
do not
necessarily state or reflect those of the United States
Government or any
agency thereof, or The Regents of the University of California.
\end{quotation}
\end{scriptsize}

\vskip 2in

\begin{center}
\begin{small}
{\it Lawrence Berkeley National Laboratory is an equal opportunity employer.}
\end{small}
\end{center}

\newpage
\renewcommand{\thepage}{\arabic{page}}
\setcounter{page}{1}

\section{Introduction}

One of the most beautiful ideas for physics beyond the Standard 
Model (SM)   
is the idea \cite{georgi} 
that the gauge groups of the Standard Model (SM)  
unify into a single gauge group, the Grand Unified Theory (GUT).
This would provide some common understanding for the diversity of 
particle content and parameters that constitute the Standard Model.
That one generation of fermions can be accommodated by a single ${\bf 16}$ of
$SO(10)$ is too remarkable to be a coincidence!
More indirect evidence for this framework is provided 
by the precision electroweak data. 
These suggest that the gauge couplings of the Standard Model 
unify at a high energy scale. 
In fact, a very good agreement with the data is 
obtained if softly-broken supersymmetry is realised   
close to the weak scale.

This naturally leads to a consideration of supersymmetric GUTs \cite{susygut}.
The scale of supersymmetric unification inferred from the data is 
$M_{GUT} \sim 2 \times 10^{16}$ GeV. Above this scale Nature may 
be described by a supersymmetric GUT.
The value of this scale given by the data does not appear to be 
directly related to any other mass scale in Nature. 
The closest scale is the reduced Planck mass, $M=1/\sqrt{8 \pi G_N}$, 
which is about a factor of 100 larger than the GUT scale.
Most attempts at supersymmetric model building remain agnostic about the
origin of the GUT scale, and simply input both the
scale and pattern of symmetry breaking
into the theory by hand. While this is technically
natural in supersymmetric
theories, it completely avoids the issues of the origin of
the GUT symmetry breaking and the small value of $M_{GUT}/ M$.
This issue is particularly relevant if the scale $M$ is representative of 
a fundamental scale of new physics.
If this is the case, then the small value of 
the supersymmetric 
Grand Unification scale compared to the Planck scale is perplexing.

Some of these issues can be addressed   
by applying some of the recent developments in the 
strong dynamics of supersymmetric gauge theories \cite{seiberg}.
In particular, the strong dynamics of an additional gauge group that 
confines with chiral symmetry breaking at a scale close to the GUT scale 
is considered.
The idea of using strong dynamics 
to generate the supersymmetric GUT scale has only recently been 
explored \cite{hc,rattazzi,yanagida}. This was first 
explored in Reference \cite{hc}, where  
a dynamically generated superpotential with a runaway behavior 
is used to generate $M_{GUT}/M$. In Reference \cite{yanagida}
  the confining
dynamics without chiral symmetry breaking is used in a novel manner 
to solve the 
doublet-triplet splitting problem. In Reference \cite{rattazzi} 
the quantum confinement with chiral symmetry breaking is used 
to generate the GUT scale. 

The idea of using strong supersymmetric dynamics to 
generate ratios of symmetry breaking scales has also been 
applied to flavour symmetries \cite{hall4, hall}. 
The first phenomenological application of 
quantum confinement with chiral symmetry breaking in this context 
is given in 
Reference \cite{hall}. 

The outline of this paper is as follows. Section 2 describes some
features that are common to the models presented in Section 3 and 4. 
Section 3 introduces 
a model with an $SU(6)$ GUT group. Section 4 introduces the 
preferred model which has an  
$SO(10)$ GUT group.

\section{Overview}

In the models presented in this paper 
an extra gauge group $G_C$ is 
introduced and assumed to become strong at a scale 
$\Lambda \sim M_{GUT}$. 
The particle content of $G_C$ is chosen so that it confines 
with chiral symmetry breaking. 
This sector of the theory will be called the `confining sector'.
By identifying the GUT group 
with a global symmetry of the 
confining sector, the composite fields of the confining sector are charged 
under the GUT group. 
For example,
in the first model presented
below, an adjoint of $SU(6)_{GUT}$ is composite. In the second model, a
symmetric and antisymmetric tensor of $SO(10)_{GUT}$ is composite.
This differs from the model of Reference 
\cite{rattazzi}, where the confining sector in that model 
does not contain particles 
charged under the GUT group.
Below the scale of confinement, some of the composite fields will
acquire vacuum expectation values (vevs) as a consequence of the
dynamics of confinement. 
In the models presented here
there is a discrete set of supersymmetric vacua. In one of 
these vacua the vevs of the composite fields break the GUT group;
this together with some superpotential 
interactions lead to a phenomenologically acceptable vacuum.
The small value of $M_{GUT}/ M_{PL}$ is then
understood as naturally arising
from the dimensional transmutation of the small gauge coupling
of $G_C$ at the Planck scale.

The simplest 
example of a supersymmetric gauge theory that exhibits confinement with 
chiral symmetry breaking is 
$SU(N)$ with $N$ flavours $Q+\overline{Q}$ and  
no superpotential \cite{seiberg}. This will be the model for the confining 
sector. It is conjectured 
that below the scale of strong dynamics, $\Lambda$,
 of the $SU(N)$ group, the 
appropriate degrees of freedom are the confined ``baryons" $B$, $\overline{B}$,
 and ``mesons" $M$
of the $SU(N)$ group, where
\begin{equation}
 M_{i}^{j} \sim \overline{Q}_a ^j Q^a _i \hbox{ }  \sim \hbox{ } 
(\overline{\Yfund}, \Yfund, \hbox{ }0)
\end{equation}
\begin{equation}
B \sim \epsilon_{a_1 \cdots a_N} Q^{a_1}_{i_1} \cdots Q^{a_N}_{i_N}
\hbox{ } \sim \hbox{ } ({\bf 1}, {\bf 1},\hbox{ } 1)
\end{equation}
\begin{equation}
\overline{B} \sim \epsilon^{a_1 \cdots a_N} \overline{Q}_{a_1} ^{i_1} 
\cdots \overline{Q}_{a_N} ^{i_N} \hbox{ } \sim \hbox{ } 
({\bf 1}, {\bf 1},\hbox{ } -1).
\end{equation} 
The charges of the baryons and mesons under the global 
$SU(N) \times SU(N) \times U(1)_{B^{\prime}}$ are indicated
in parantheses. The space of supersymmetric vacua for the baryons
and mesons is described by \cite{seiberg} 
\begin{equation}
 \det M -B \overline{B}= \Lambda ^{2N}.
\label{qc}
\end{equation}
The left-hand-side of this equation vanishes at the classical level 
as a consequence of the Bose statistics of the superfields $Q$ and 
$\overline{Q}$. Quantum corrections result in a non-vanishing value
for the right-hand-side. 
The important point is that along the supersymmetric vacua, some 
of the confined fields necessarily acquire vevs, 
breaking the 
global symmetry 
down to a subgoup. This conjecture
satisfies two nontrivial consistency tests \cite{seiberg}: 
holomorphic decoupling 
of one flavour; and t'Hooft anomaly matching of the unbroken global symmetries.
 
In this paper a diagonal subgroup of the global symmetry of the 
confining sector is gauged and identified with the GUT group.
The mesons of the confining sector therefore transform under the 
GUT group. I will make the dynamical assumption that
weakly gauging a global symmetry of the confining sector 
does not affect the confining dynamics of $G_C$,
and does not ruin the quantum modification with chiral
symmetry breaking. This is a reasonable assumption since
the GUT group is weakly gauged at the scale
$\Lambda \sim M_{GUT} \sim
$2$\times$10$^{16}$ GeV.  

Perhaps the most difficult problem in GUT model building is the origin of
the doublet-triplet mass splitting. The excellent agreement between the
measured and theoretically predicted value of $\sin^2 \theta_W$
assumes that the particle content below the unification scale
contains the (supersymmetric) 
SM chiral matter content plus two electroweak Higgs doublets.
In a minimal $SU(5)$ GUT, the Higgs fields fit 
into a ${\bf 5}$ and $\overline{{\bf 5}}$ of $SU(5)$.
The presence of the remaining particle content of these 
representations-the two coloured Higgs triplets-
much further than a few decades below the GUT scale completely
ruins this agreement. More generally, requiring that there exists one large 
split $SU(5)$ representation is a strong 
constraint on 
model building. 
The models presented in this paper implement two known solutions to
this problem: the Higgs as ``pseudo-Goldstone bosons" \cite{su6} and the
``Dimopoulos-Wilzcek" \cite{dw} missing vevs mechanism. The latter solution is
implemented in an $SO(10)$ GUT gauge group, whereas the former is based
upon an $SU(6)$ GUT group.    

In the models presented here the quantum confinement is 
therefore not directly responsible for the doublet-triplet splitting.  
The structure outlined above must be supplemented with a 
non-vanishing superpotential in order to implement the doublet-triplet 
splitting. A non-vanishing superpotential must be added in 
any case: 
a generic point on the quantum modified constraint breaks $
SU(N) \times U(1)_{B^{\prime}}$ down 
to $U(1)^{N-1}$. This provides too much symmetry breaking. 
A point that only breaks to a larger subgroup is therefore 
an enhanced symmetry point, corresponding to a particular choice of the 
vevs of $M$ and $B$. 
At the enhanced symmetry point, 
there are many massless particles in addition to the 
Nambu-Goldstone multiplets. These correspond to the would-be Goldstone 
bosons of the more generic symmetry breaking pattern, and 
at the enhanced symmetry point, 
transform as adjoints under the unbroken gauge group.
These particles 
must acquire 
masses from additional superpotential interactions. 

It is then a concern whether the presence of this superpotential
might destabilise the confinement and chiral symmetry breaking.
The form of the superpotential
for the fundamental fields of the group $G_C$, $Q$, $\overline{Q}$, and 
any fields $\psi_M$ not charged under $G_C$, in the two models presented here is 
\begin{equation}
W=W_C(Q,\overline{Q}, \psi_M)+W_M(\psi_M).
\end{equation}
The superpotential $W_C$ involving the confining
fields will by {\it fiat} contain only 
non-renormalizable operators, suppressed by a scale 
assumed to be either 
the Planck mass or reduced Planck mass. 
If 
confinement occurs, 
the coefficient $c$ of an operator with mass dimension $d$ 
in the low-energy theory 
that arose from an operator with $N$ $(\overline{Q} Q)$s in the 
high energy theory   
is expected to be 
\begin{equation}
 c \sim \lambda \times \Lambda ^N /M^{N-d},
\end{equation}	
where $\lambda$ is a constant that is expected to be of 
order unity. 
For the models considered below, $d=-1,0$ or 
1, $N$ is 1 or 2, and $N-d$ is postive. 
Since these coefficients are suppressed by powers of 
$\Lambda /M$,  
the presence of these terms in the superpotential is a small 
perturbation to the quantum confinement. It is then reasonable 
to expect that these operators do not destroy the quantum confinement with 
chiral symmetry breaking. This assumption will be made   
for the remainder of the paper. 

In the usual GUT model building framework, the unification 
of the gauge couplings can be significantly affected by 
the presence of $M^{-1}$ 
suppressed operators \cite{hall2}. 
In an $SU(5)$ model, for example, 
the gauge field-strength tensor $F$ can have non-renormalizable interactions 
with an adjoint $\Sigma$. The 
operator $c \Sigma F F /4 M$ results in a tree-level 
relative shift of the gauge couplings $1/ g^2 _i$ 
that is approximately $ cM_{GUT}/M$. This translates into a shift in 
the low-energy value of $\sin \theta ^2 _W$ that for $M/ M_{GUT}=20$ is 
$\Delta \sin \theta ^2 _W (M_Z)\sim \pm \hbox{few}\times c \times 10^{-3}$. 
In the GUT models presented in this paper,
some of the higher dimensional 
representations are composite. 
For the composite fields, 
the gravitational smearing operator arises from a higher 
dimension operator in the fundamental theory. The coefficient of 
this operator below the confinement scale then  
contains an additional suppression of $\Lambda /M$. 
This extra factor completely suppresses the smearing effect 
unless the coefficient of the operator in the fundamental theory is 
unnaturally large-of $O(M/M_{GUT})$-and $M_{GUT} /M$ is $\sim 1/20$.
Non-composite higher dimensional fields can contribute to the 
gravitational smearing. In the $SO(10)$ model, it turns out that these 
contributions are completely negligible.

I conclude this Section with a discussion of 
some technical issues that occur 
throughout the paper. Implicit in the discussion that follows 
will be the assumptions
that (global) supersymmetry is unbroken, 
and that the non-trivial Kahler
potential has a strictly positive definite Kahler metric \cite{hall}.  

To find supersymmetric minima I will look for solutions to the $F-$flatness
equations $0=F=\partial_{\phi_i} W$ for the confined and $\psi _M$ fields. 
This is rather naive, since
the vevs of the fields will typically
be $O(\Lambda)$ and
the Kahler potential
is non-calculable for these field values. 
It is not clear then that the ``baryons" and
``mesons"
are the correct degrees of freedom. For the purposes of determining
the existence of supersymmetric vacua with a particular pattern of
symmetry breaking, however, the last assumption of the previous 
paragraph is suffucient \cite{hall}.
With these assumptions, a supersymmetric vacuum found using
a trivial Kahler potential will remain supersymmetric for the non-trivial
Kahler potential. 

The spectrum of the particle 
masses is also important for phenomenology. For this, 
knowledge of the Kahler potential is required. Despite the  
absence of this information,  
a few important points about the mass spectrum can be extracted from 
the superpotential \cite{hall}.
For example, a particle that is massless (zero eigenvector of $F_{i,k}$) 
in the case of a canonical Kahler potential
for the confined fields will remain 
massless in the case of a non-trivial Kahler potential. 
Similarly, a massive particle in the trivial Kahler potential 
will remain massive for a non-trivial  
Kahler potential.  
So I will use the mass spectrum computed by assuming a  
trivial Kahler potential to check that the 
superpotential with a non-trivial  
Kahler potential results in 
superheavy masses to all the particles that should have 
superheavy masses.

In the models presented here, the 
superpotential interactions that involve the confining fields occur
from higher dimension operators, so that after confinement
the superpotential coupling of those operators is 
$\tilde{\lambda} \sim \lambda (\Lambda /M)^n \ll \lambda$, with
$\lambda \sim O(1)$. 
Particles that acquire their mass from these operators will then have 
masses somewhat below the GUT scale. These masses remain uncalculable though,
since they should be computed at a scale that is comparable to the 
{\it vev} that is generating the mass, which in this case is $O(\Lambda)$.

The one-loop prediction for $\sin ^2 \theta _W$ is modified by the 
presence of these light states below the GUT scale since 
they do not in general form complete $SU(5)$
representations. 
An attempt at quantifying this correction is made 
by assuming that
the naive calculation-i.e. assuming a canonical
Kahler potential-of the spectrum 
gives the correct mass spectrum to within a few factors
of unity, and further, that the correction to $\sin ^2 \theta _W$
from particles with masses much smaller than the
confinement scale is well-approximated by the usual one-loop computation.
The corrections from particles with masses near the confinement scale are not
calculable and not discussed.

Finally, in the two models presented here 
certain operators 
allowed by the gauge symmetries of the theory must be absent from 
the superpotential in order 
not to ruin the doublet-triplet splitting mechanisms. 
All the dangerous operators cannot 
be forbidden by any global symmetries, since some of them will have the same 
quantum numbers as other operators that are required to be 
present in the superpotential. 
If these models were only the effective theory of some more 
fundamental field theory, then the dangerous 
operators could perhaps be generated at the 
tree-level by integrating
out some heavy particles at the scale $M$. In this case however, 
the full theory above the 
Planck scale is not known and probably not a field theory. It is then 
possible 
that  
the full theory could be 
responsible for the absence of these dangerous operators, even though from 
the low-energy theory they cannot be forbidden by any symmetries. 

\section{$SU(6) \times SU(6)$}

The gauge group is $SU(6)_C \times SU(6)_{GUT}$ where 
one factor of $SU(6)$ is the
confining group $G_C$, and the other factor is the SM unified gauge 
group. 
I introduce six flavours, $Q+\overline{Q}$ of $SU(6)$ 
that are also charged under 
the $SU(6)_{GUT}$. I further introduce two Higgs fields $H$, $\overline{H}$, 
and an adjoint $\Sigma_N$ that 
are charged under only the $SU(6)_{GUT}$. The particle content 
under $SU(6)_C \times SU(6)_{GUT}$ is then
\begin{eqnarray*}
Q & \hbox{ }\sim \hbox{ } & ({\bf 6},\hbox{ } {\bf \overline{6}} ), \\
\overline{Q} & \hbox{ } \sim \hbox{ } & 
({\bf \overline{6}},\hbox{ } {\bf 6}), \\
H & \hbox{ } \sim \hbox{ } &  ({\bf 1},\hbox{ } {\bf 6}), \\
\overline{H} & \hbox{ } \sim \hbox{ } & 
({\bf 1},\hbox{ } {\bf \overline{6}}), \\
\Sigma_N & \hbox{ } \sim \hbox{ } &  ({\bf 1},\hbox{ } {\bf 35}) . 
\end{eqnarray*}
I assume that the $SU(6)_C$ group confines at a scale $\Lambda \sim M_{GUT}$
with a quantum modified constraint. In this case the confined ``meson" 
$M_i ^j \sim \overline{Q} ^a _i Q^j _a \sim {\bf 35 +1}$ under
 the $SU(6)$ GUT 
symmetry. The ``baryons" $B \sim \epsilon Q^6$ and $\overline{B} \sim \epsilon
 \overline{Q} ^6$ are singlets under the $SU(6)_{GUT}$ group.
No gauge singlets are required in the fundamental theory.

The superpotential in terms of the fundamental fields is chosen to be 
\begin{equation}
W_0 = \frac{1}{2}\lambda_1 \hbox{tr}(Q \overline{Q})^2/ M +   
\lambda_3 H \overline{H} \hbox{tr}(Q \overline{Q})/M + 
\lambda \hbox{tr} (\Sigma^2 _N Q \overline{Q}) /M + \bar{g} (H \overline{H}) 
\hbox{tr} \Sigma_N ^2 /M. 
\end{equation}
The scale $M$ is assumed to be the reduced Planck mass $\sim $2$\times 
$10$^{18}$ GeV. The trace sums over the $SU(6)_{GUT}$ indices. 
All the dimensionless parameters are assumed to be of order unity.
This superpotential is the most minimal, in the sense that (as shown below) 
it successfully implements in the phenomenologically preferred 
vacuum the doublet-triplet splitting and gives 
GUT scale masses to all the other particles. A more general superpotential 
is allowed provided that: (1) Only non-renormalizable operators 
involving $Q$, $\overline{Q}$ are allowed. This is guaranteed by the 
$SU(6)_C$ gauge symmetry if mass terms are forbidden. 
(2) To keep the Higgs doublets light, the superpotential 
that only involves the ${\bf 35}$s and the $H$, $\bar{H}$ fields must preserve 
a $SU(6) \times SU(6)$ global symmetry. The operators 
$\overline{H} (\overline{Q} Q ) ^{n} H$ and $\overline{H} \Sigma ^n _N H$, 
for example, must be absent. 
(3) Supersymmetry is not spontaneously broken. 
For example, a
 cubic term for $\Sigma_N$ which
satisfies (1), (2) and (3) is allowed,
but not required.

After confinement occurs, the superpotential written in terms of the confined
fields $\Sigma \sim {\bf 35}$ and $\sigma \sim {\bf 1}$ , i.e.  
$\overline{Q} ^a _i Q^j _a \sim \Lambda \Sigma _i ^j + \Lambda 
\sigma \delta _i ^j/ \sqrt{6}$, is
\begin{eqnarray*}
W_0&=& {\cal A} 
\left(\det (\Sigma + \sigma/ \sqrt{6}) -B \overline{B} -\Lambda^6 \right)
+\frac{1}{2}\tilde{\lambda}_1 \Lambda \hbox{tr} \Sigma^2 
+\frac{1}{2}\tilde{\lambda}_2 \Lambda
\sigma^2 \\ 
 & & +\tilde{\lambda}_4 \hbox{tr} \Sigma^2 _N \Sigma + 
\tilde{\lambda}_5 \sigma \hbox{tr} \Sigma^2 _N+ 
(H \overline{H}) (\tilde{\lambda}_3 \sigma+ \bar{g} \hbox{tr} \Sigma^2 _N /M) . 
\end{eqnarray*}
I expect that 
\begin{equation}
 \tilde{\lambda}_{1,2} \hbox{ } \sim \hbox{ }\lambda_{1} \Lambda / M, 
\hbox{ } \hbox{ } \tilde{\lambda}_3 \hbox{ } \sim \hbox{ } \lambda_3 
\Lambda / M,  \hbox{ } \tilde{\lambda}_{4,5} \sim \lambda \Lambda /M
\end{equation} 
as an estimate of the size of the couplings in the confined description.
The quantum modified constraint has been added using a Lagrange 
multiplier ${\cal A}$.
This superpotential contains all 
 the non-perturbative (superpotential) information from the                   
strong $SU(6)_C$ dynamics. 
It is interesting that 
in this case a term in the superpotential for $Q \overline{Q}$ that 
generates a cubic term $\hbox{tr} \Sigma ^3$ is not required. 
In most  
supersymmetric GUT models, the cubic term  
is required to obtain a non-trivial vacuum. In this case, it is the 
interaction ${\cal A} \det (\Sigma + \sigma)$ from 
the quantum modified constraint that balances the mass terms to 
obtain a non-trivial supersymmetric vacuum. 

The $F-$flatness equations are  
\begin{equation}
\det (\Sigma+\sigma/ \sqrt{6})-B \overline{B}=\Lambda ^6,
\end{equation}
\begin{equation}
0=F_{\overline{B}}={\cal A} B, \hbox{ } 0=F_{B}={\cal A} \overline{B} ,
\end{equation}
\begin{equation}
0=F_{\overline{H}}=(\tilde{\lambda}_3 \sigma+ g \hbox{tr} \Sigma^2 _N /M) H,
\label{H}
\end{equation}
\begin{equation}
0=F_{H}=(\tilde{\lambda}_3 \sigma + g \hbox{tr} \Sigma^2 _N /M) \overline{H},
\end{equation}
\begin{equation}
0=F_{\sigma}=\tilde{\lambda}_3 H\overline{H}
+\tilde{\lambda}_2 \Lambda \sigma 
+ \frac{{\cal A}}{\sqrt{6}} \det( \Sigma+\sigma / \sqrt{6})
\hbox{tr}(\Sigma+\sigma / \sqrt{6})^{-1}+\tilde{\lambda}_5 
\hbox{tr}{\Sigma^2 _N},
\label{ls}
\end{equation}
\begin{equation}
0=F_{\Sigma}=\tilde{\lambda}_1 \Lambda \Sigma
+{\cal A} \det (\Sigma+\sigma / \sqrt{6})((\Sigma+\sigma / \sqrt{6})^{-1}-
\frac{1}{6} \hbox{tr}(\Sigma+\sigma / \sqrt{6})^{-1})+ 
\tilde{\lambda}_4 ( \Sigma^2 _N -\frac{1}{6} \hbox{tr} \Sigma^2 _N),
\label{s}
\end{equation}
\begin{equation}
0=F_{\Sigma_N}=\tilde{\lambda}_4 ( \Sigma _N \Sigma - 
\frac{1}{6} \hbox{tr} \Sigma_N \Sigma)+ \tilde{\lambda}_5 \sigma \Sigma_N 
+ g (H \overline{H}) \Sigma_N /M.
\end{equation}
In addition to 
the phenomenologically preferred vacuum, these equations include 
other discrete solutions.  
In some of these solutions $SU(6)_{GUT}$ is unbroken. For example, 
a solution with
$\sigma $ and ${\cal A}$ non-zero, and all other vevs equal to zero,
exists.
So although the preferred vacuum is discrete, I must assume that
it was selected in the early history of the universe.
This could occur if, for example,
the preferred vacuum is a global minimum of the scalar potential after
supersymmetry breaking effects are
included.

To break $SU(6)$ down to the SM gauge group, I look for vevs of the form
\footnote{$H=\overline{H}$ 
is required by $SU(6)_{GUT}$
 $D-$flatness.}
\begin{equation}
H=\overline{H}=v_H \left(\begin{array}{c}
1 \\ 0 \\ 0 \\ 0 \\ 0 \\ 0 \end{array} \right) \hbox{ },  
\hbox{ } \Sigma(\Sigma_N)= v_{\Sigma}(v_N) \left(\begin{array}{cccccc}
1 & & & & & \\
 & 1 & & & & \\ 
 &   &1& & & \\
 &   & &1& & \\
 &   & & &-2& \\
 &   & & &  & -2 \end{array} \right).
\label{vacuum}
\end{equation}
The vevs ${\cal A}$, $\sigma$, 
$v_{\Sigma}$, $v_N$ and $v_H$ are the solution to
\begin{equation}
0=(\tilde{\lambda}_3 \sigma + 12 g v^2_N /M)v_H ,
\end{equation}
\begin{equation}
0=(\tilde{\lambda}_5 \sigma + g v_H^2/M - \tilde{\lambda}_4 v_{\Sigma})v_N, 
\end{equation}
\begin{equation}
0= \frac{1}{3} {\cal A} K (a-b) + \tilde{\lambda}_1 \Lambda v_{\Sigma} 
- \tilde{\lambda}_4 v^2 _N ,
\end{equation}
\begin{equation}
0= 2 \frac{{\cal A} K}{\sqrt{6}} (2a +b) + \tilde{\lambda}_2 \Lambda 
\sigma + \tilde{\lambda}_3 v^2 _H + 12 \tilde{\lambda}_5 v^2 _{N},
\end{equation}
and for ${\cal A} \neq 0$, $\det (\Sigma+ \sigma/ \sqrt{6}) =\Lambda^6$.
The quantities $a$, $b$ and $K$ are defined to be 
$a^{-1} \equiv v_{\Sigma} + \sigma / \sqrt{6} $, $b^{-1} \equiv -2 v_{\Sigma}
+ \sigma / \sqrt{6} $ and $K \equiv \det ( \Sigma + \sigma / \sqrt{6}) =
a^{-4} b^{-2}$. 
In Appendix A it is demonstrated that a discrete solution 
exists with  ${\cal A} \sim (\Lambda/ M) \Lambda ^{-3}$ and with all 
vevs non-zero and of $O(\Lambda)$. 

  
This vacuum implements the Higgs as ``pseudo-Goldstone bosons'' solution
to the doublet-triplet splitting problem \cite{su6}. This mechanism 
is now briefly 
described. Firstly, the scalar 
potential for $H$, 
$\overline{H}$ and $\Sigma$, $\Sigma_N$  
has a $U(6) \times SU(6)$ global symmetry. 
The $U(6)$ acts on $H$ and $\overline{H}$, whereas the $SU(6)$ symmetry
acts on $\Sigma$ and $\Sigma_N$. For the vacuum in Equation \ref{vacuum}, 
the global $U(6) \times SU(6)$ symmetry is broken to $[SU(5)] \times
[SU(4) \times SU(2) \times U(1)]$ by the vevs of $H$, $\Sigma$ and 
$\Sigma_N$. 
The unbroken 
gauge group is then $SU(3)_C \times SU(2) \times U(1)_{Y}$. The 
breaking of the gauge symmetry results in 23 Nambu-Goldstone boson 
multiplets; the breaking of the $SU(6) \times U(6)$ results in 
27 Goldstone boson multiplets.   
So all but 4 of the Goldstone bosons
acquire mass of $O(M_{GUT})$ from the super-Higgs mechanism.

To see that these four pseudo-Goldstone bosons carry the quantum number  
charges of two electroweak doublets, first note that under 
$SU(4) \times SU(2)$, ${\bf 35}=({\bf 4},{\bf 2})+(\overline{{\bf 4}},{\bf2})
+({\bf 15},{\bf 1})+({\bf 1},{\bf 3})+({\bf 1}, {\bf 1})$. Inspecting the 
vevs of $\Sigma$ and $\Sigma_N$, the combination 
$\tilde{v} _{\Sigma} \tilde{\Sigma} \equiv  v_{\Sigma} \Sigma 
+ v_N \Sigma _N $ of the 
fields $({\bf 4},{\bf 2})$, and of the fields $(\overline{{\bf 4}},{\bf2})$,
 in $\Sigma$ and 
$\Sigma_N$  
are the Goldstone bosons of the breaking of one global $SU(6)$ symmetry. 
Since $SU(3)_C$ is embedded in 
$SU(4)$, these Goldstone bosons contain two electroweak doublets. The 
Goldstone bosons of the $SU(6) \rightarrow SU(5)$ breaking are 
${\bf 5}+\overline{{\bf 5}}+{\bf 1}$ of $SU(5)$, and also contain two 
electroweak doublets. The combination $ 3 \tilde{v}
_{\Sigma} \tilde{\Sigma}
+v_{H} H$ 
of electroweak Higgs doublets 
are the fields eaten by the super-Higgs mechanism. The orthogonal 
combination remain massless and are the two Higgs doublets of the SM. 
The non-renormalization theorems of supersymmetry guarantee that 
these fields remain massless to all orders in perturbation theory.

The fields in the adjoint 
$({\bf 15},{\bf1})$
and $({\bf 1},{\bf 3})$ of both $\Sigma$ and $\Sigma_N$, as well as 
the remaining combination of $({\bf 4},{\bf 2})$, and of 
$(\overline{{\bf 4}},{\bf 2})$, in 
$\Sigma$ and $\Sigma_N$ orthogonal to $\tilde{\Sigma}$, 
do not correspond to any broken 
generators and must acquire their masses 
from the superpotential interactions. It is conveinent to express 
the $SU(5)$ or SM charge assignments of this particle content:
one complete ${\bf 24}$ and ${\bf 5}+
{\bf \overline{5}}$ of $SU(5)$; 4 singlets; and one
$({\bf 8}, {\bf 1}, 0)+ ({\bf 1}, {\bf 3},0)+({\bf 3},{\bf 1}, -1/3)+
({\overline{\bf 3}}, {\bf 1}, 1/3)$.
A naive estimate for the masses of the physical fields 
is obtained by computing the fermion
mass matrix assuming a canonical Kahler potential. The results 
are presented in Appendix A, and are summarized here. 
All the fields have a mass $m \sim \Lambda^2 /M$, a consequence of 
the suppression of the superpotential couplings for the confined theory.

These light fields affect the unification of the gauge couplings 
and may in principle also mediate proton decay. 
I first discuss the corrections to 
$\sin ^2 \theta _W$. These corrections occur from two sources. 
There could be large threshold corrections
from the strong dynamics occuring at $\Lambda$. These 
are non-calculable
and will not be considered. 
The other is from the light states 
$({\bf 8},{\bf 1},0)$, $({\bf 3},{\bf 1},0)$, 
$(\overline{ {\bf 3}}, {\bf 1},1/3)$ 
 and $({\bf 3}, {\bf 1},-1/3)$ which have a mass 
$m \sim 
\Lambda ^2 / M$ . 
The correction to $\sin ^2 \theta _W$ from 
these light states, using a naive one-loop running approximation 
from $M_{GUT}$ to their 
masses is 
\begin{equation}
\Delta \sin ^2 \theta _W =- \frac{\alpha_{em}}{5 \pi} \ln M_{GUT} /m 
\sim -0.003 \times 
\frac{\ln ( M_{GUT} / m)}{\ln 200} .
\label{sin}
\end{equation}
The 
reason\footnote{The author thanks N. Arkani-Hamed for this 
observation.} for the small correction
is that the shift in 
$\sin ^2 \theta _W$ is dominated by the 
light $(\overline{{\bf 3}},{\bf 1}, 1/3)$ and 
 $({\bf 3},{\bf 1},-1/3)$ states. This is 
 because the shift from the $({\bf 8}, {\bf 1})$ 
and $({\bf 1}, {\bf 3})$ states almost cancel. Recall that a 
sufficient condition for the prediction for $\sin ^2 \theta _W$ to be 
unchanged by the presence of some extra matter at a scale $m$ 
is that $(\delta b_3 - \delta b_2)/ (\delta b_2- \delta b_1) 
=(b_3 -b_2)/(b_2 -b_1)$, 
independent of $m$. For an adjoint of $SU(3)$ and 
$SU(2)$ , $\delta b_3=3$, $\delta b_2 =2$ and $\delta b_1=0$. In this 
case the LHS of this condition is 1 and the RHS is $\frac{5}{7} \times 2$, 
which is close to 1. 
The other light states form approximate complete $SU(5)$ representations
and do not significantly affect the gauge-coupling unification.
The theoretical prediction without the light fields, 
$ \sin ^2 \theta _W \sim .233 \pm O( 10^{-3})$ \cite{sin},
is a little larger than the measured value of 0.231\cite{pdg}. 
The effect of these 
light states is to shift the prediction in the correct direction.
The uncertainty in the uncalculable corrections to $\sin ^2 \theta_W$, 
however,  
are probably of the same order, with an unknown sign. 

I next discuss the problem of forbidding operators of the form 
${\cal O}_n \sim 
\overline{H} (Q \overline{Q} )^n H$. These operators explicitly break 
the $U(6) \times SU(6)$ symmetry of the scalar potential. 
Consequently, if these operators are present they 
could give too
large of a mass to the electroweak Higgs doublets. 
In this model, the term $\overline{H} H \hbox{tr} \overline{Q} Q$ occurs
in the superpotential. Any symmetry that allows this term 
also allows the term 
$\overline{H} 
(\overline{Q} Q )H$ in the superpotential. This operator ruins the 
doublet-triplet splitting, so I must assume that this term is absent. 
Higher dimensional operators must also be forbidden. Since 
the confinement introduces additional suppressions of 
$O(\Lambda^n /M^n)$, only a few of the first higher dimensional 
operators must be absent. More concretely, if I require 
that ${\cal O}_n$
not result in a mass for the Higgs superfields that is larger 
than a TeV and assume that $\Lambda / M_{PL} \sim 1/200$, then
only the first three ($n=$1,2 and 3) higher dimensional operators 
must be forbidden. Operators of the type $\overline{H} (\Sigma_N)^n H$ 
are also dangerous and must be absent. 

At this point it is probably not clear what role the extra adjoint 
plays in this model. In fact, this field is not 
needed to obtain an acceptable 
spectrum for the massive fields. It is introduced instead to obtain a large 
top quark Yukawa coupling. In order for the top quark not to have 
an irrelevant Yukawa coupling, it is necessary that the Yukawa 
interactions between the top quark and the Higgs doublet  
explicitly break the global $SU(6) \times U(6)$ 
symmetry. The top quark must therefore couple to both $H$ and 
$\Sigma$. If $\Sigma $ is composite, then such a coupling cannot 
be of order unity; rather, it will be suppressed by $\Lambda /M$. 
The top quark must therefore interact 
with a fundamental $\Sigma$. 

The large top quark Yukawa coupling arises from considering the following 
embedding of the SM chiral fields \cite{hall3} . 
The chiral matter content is one ${\bf 20}$, $3 \times {\bf 15}$ and 
$6 \times \overline{\bf 6}$. The $SU(5)$ decomposition of these fields is, 
${\bf 20}={\bf 10}+\overline{\bf 10}$, ${\bf 15}={\bf 10}+{\bf 5}$ and 
$\overline{\bf 6}=\overline{\bf 5}+{\bf 1}$. The three $\overline{{\bf 5}}$s 
of the SM are contained in three of the $\overline{\bf 6}$s, and the 
other 3, call them $\overline{\bf 6} ^{\prime}$, acquire mass at the 
GUT scale. The first two generation ${\bf 10}$s are contained in two 
of the ${\bf 15}$s, and the third generation ${\bf 10}$ is a linear 
combination of the ${\bf 10}$ in the ${\bf 20}$ and the 
${\bf 10}$ in the remaining ${\bf 15} \equiv {\bf 15}_3$. 
This spectrum is obtained from the superpotential \cite{hall3}
\begin{equation}
W_{top}= \lambda {\bf 20} \Sigma_N {\bf 20} 
+ \lambda^{\prime} {\bf 20} H {\bf 15}_3 +
\lambda^{\prime \prime }_{ij} 
\overline{H} {\bf 15}_i \overline{{\bf 6}}^{\prime} _j.
\label{top}
\end{equation} 
The vev of $\overline{H}$ gives GUT-sized Dirac masses to the  
${\bf 5}$ and $\overline{{\bf 5}}$ fields in the 3 ${\bf 15}$s and 
3 $\overline{\bf 6} ^{\prime}$s. From the vevs of $\Sigma_N $ 
and $H$, a linear combination of the ${\bf 10}$ in the ${\bf 20}$ 
and the ${\bf 10}$ in ${\bf 15}_3$ acquires a GUT-sized Dirac mass with 
the $\overline{\bf 10}$ 
in the ${\bf 20}$. The orthogonal combination is the third generation 
${\bf 10}$ and remains massless. In sum, this superpotential
 leaves 3 $({\bf 10} + \overline{{\bf 5}})$s massless.  
The large top quark Yukawa coupling arises from the first two interactions. 

The $({\bf 3},{\bf 1}, 1/3)$ and $(\overline{\bf 3},{\bf 1},-1/3)$
fields have a Dirac mass somewhat below the GUT scale.  
Whether they may mediate proton 
decay at too large of a rate is then a concern. 
Since the top quark couples to these fields through the 
${\bf 20} \Sigma_N {\bf 20}$ interaction, it naively appears that 
a 
dangerous proton decay operator
is generated by integrating out these heavy fields, and then 
rotating the top quark to the mass basis. For this operator to be generated, 
however, a coupling of 
$\Sigma_N$ or $\Sigma$ to a $\overline{{\bf 5}}$ of $SU(5)$ 
($\overline{{\bf 6}} $ of $SU(6)$)
is required. Such a coupling is not present in 
the superpotential of Equation \ref{top}. So 
this issue depends crucially on the origin of the other 
fermion masses. For example, if all the fermion masses arise from interactions
with $H$ and $\overline{H}$, then a dangerous proton decay 
operator is not 
generated by the exchange of 
these states \cite{hall3}.

An upper bound on $M$ is determined by the value of the Landau pole 
of the $SU(6)_{GUT}$ gauge coupling. 
The $SU(6)$ coupling at the scale $M$ is  
then 
\begin{equation}
\alpha ^{-1}_{SU(6)_{GUT}} (M)=24-\frac{43}{11 \pi} \ln \Lambda / m_I 
- \frac{7}{2 \pi} \ln M / \Lambda . 
\end{equation}
The first logarithm is the contribution to the GUT gauge coupling at the 
GUT scale from the particle content 
with mass $m$; the second logarithm 
is the contribution of the full $SU(6)$ particle content to the running 
of the gauge coupling above $\Lambda$.
Inserting $m \sim \Lambda ^2 /M$ and 
requiring that $\alpha ^{-1}_{GUT} (M) \geq 1$ implies 
$\ln M / \Lambda \leq 10$. 

\section{ $SU(10) \times SO(10)$}

The gauge group is $SU(10)_C \times SO(10)$. The $SU(10)_C$ group 
is the confining 
gauge group, and the Grand Unified group is $SO(10)$. 
The particle content is
\begin{eqnarray*}
Q & \hbox{ } \sim \hbox{ } & ( {\bf 10}, \hbox{ } {\bf 10}), \\
\overline{Q} & \hbox{ } \sim \hbox{ } &
 (\overline{{\bf 10}}, \hbox{ } {\bf 10}), \\
A & \hbox{ } \sim \hbox{ } & ({\bf 1}, \hbox{ } {\bf 45}), \\
{\bf 16} & \hbox{ } \sim \hbox{ } & ({\bf 1}, \hbox{ } {\bf 16}), \\
\overline{{\bf 16}} & \hbox{ } \sim \hbox{ } & ({\bf 1}, \hbox{ } 
\overline{{\bf 16}} ), \\
T_1 & \hbox{ } \sim \hbox{ } & ({\bf 1}, \hbox{ } {\bf 10}), \\
T_2 & \hbox{ } \sim \hbox{ } & ({\bf 1}, \hbox{ } {\bf 10}).
\end{eqnarray*}
This particle content is rather economical as it requires only 
one higher dimensional representation, an adjoint, and no 
gauge singlets. 
I assume that the $SU(10)_C$ group confines at a scale $\Lambda \sim M_{GUT}$
with a quantum modified constraint. In this case the confined ``meson"
$M_i ^j \sim \overline{Q} ^a _i Q^j _a \sim {\bf 45 +54+1}$ under
 the $SO(10)$ GUT
symmetry. I label $S \sim {\bf 54}$,  
$A^{\prime \prime} \sim {\bf 45}$ and $\sigma \sim {\bf 1}$. 
The ``baryons" $B \sim \epsilon Q^6$ and $\overline{B} \sim \epsilon
 \overline{Q} ^6$ are singlets under the $SO(10)_{GUT}$ group. 

The superpotential in the fundamental theory is chosen to be 
\begin{eqnarray}
W &= & \lambda_1 T_1 A T_2 + \lambda_2 T_2 (\overline{Q} Q) T_2 /M +
\lambda_3 \overline{{\bf 16}} (\overline{Q} Q) \Sigma {\bf 16} /M 
+ \lambda_4 \overline{{\bf 16}} {\bf 16} \hbox{tr} (\overline{Q} Q) /M \\ 
\nonumber
& & + 
\lambda_5 \hbox{tr} (\overline{Q} Q)^2 /M + 
\lambda_7 A ^2 
(\overline{Q} Q) /M + \lambda_{11} 
\overline{{\bf 16}}(Q \overline{Q} )_{AS} A 
\Sigma {\bf 16} / M^2,
\end{eqnarray} 
where $\Sigma_{ij}= [\Gamma_j, \Gamma_i]/4 i$ are the  
generators of $SO(10)$ in the spinorial representation. 
The subscript ``AS'' indicates 
that 
only the anti-symmetric contribution of $Q \overline{Q}$ is allowed to 
be present; the symmetric contribution spoils the doublet-triplet splitting.
It is technically natural for only the anti-symmetric contribution to be 
present; the full theory above the Planck scale must be responsible for the 
absence of the symmetric operator.
The operators $T_1 (\overline{Q} Q)^n T_1$ must also be absent.

The renormalizable and $M^{-1}$ suppressed operators appearing in 
$W$ are all required:
(i) the operators $\propto$ $\lambda_1$, $\lambda_2$ are required 
for the doublet-triplet splitting; (ii) the operator $\propto$ 
$\lambda_7$ arranges the vev of $A$ to be 
in the ``Dimopoulos-Wilzcek'' form \cite{dw}, required to perform the
doublet-triplet splitting;
(iii) the operators $\propto$ $\lambda_3$ and $\lambda_4$ are 
necessary to break the rank of the group; (iv) the operator $\propto$ 
 $\lambda_5$ is necessary to fix all the vevs. The operator $\propto$ 
$\lambda_{11}$ is required to give mass to some fields charged under the SM. 
This point is discussed more later. The choice for this operator is not unique; 
other operators that are linear in $A^2$ are possible, but they are higher 
dimensional. It is non-trivial that with this choice for $W$, the low-energy 
particle content only contains the SM fields and their superpartners. 
  
After confinement occurs the superpotential is 
\begin{eqnarray}
W& = & W_{H} + W_{DW} + W_{mix},  
\end{eqnarray}
with
\begin{eqnarray}
W_{H} & = & \lambda_1 T_1 A T_2 + \tilde{\lambda}_2 T_2 S T_2 
+ \tilde{\lambda}_3  \sigma T_2  T_2, \\ 
W_{DW} & = &  \frac{1}{2} \tilde{\lambda}_{9}A^2 S 
+\frac{1}{2} \tilde{\lambda}_{10} \sigma A^2, 
\\
W_{mix} & = & 
{\cal A}\left(\det (S+A^{\prime \prime}+ \sigma / \sqrt{10}) 
-B \overline{B} -\Lambda^{10}\right) 
+\frac{1}{2} \tilde{\lambda}_5 \Lambda \sigma^2 +
\frac{1}{2} \tilde{\lambda}_6 \Lambda S^2+
\frac{1}{2} \tilde{\lambda}_7 \Lambda {A^{\prime \prime}}^2
\nonumber \\ 
& & 
+ \tilde{\lambda}_{16} \sigma \overline{{\bf 16}} {\bf 16} 
+ \tilde{\lambda}_4 A^{\prime \prime} _{ij} 
\overline{{\bf 16}} \Sigma_{ij} {\bf 16} 
+ \tilde{\lambda}_{11}  (A A^{\prime \prime})_{ij} 
\overline{{\bf 16}} \Sigma_{ij} 
{\bf  16} /M.
\end{eqnarray}  
The naive expectation for the couplings is 
$\tilde{\lambda}_{2,3} \sim \lambda_2 \Lambda /M$,  
$\tilde{\lambda}_{4} \sim \lambda_3 \Lambda /M$, 
$\tilde{\lambda}_{16} \sim \lambda_4 \Lambda /M$,
$\tilde{\lambda}_{5,6,7} \sim \lambda_5 \Lambda /M$, 
$\tilde{\lambda}_{9,10} \sim \lambda_7 \Lambda /M$, and 
$\tilde{\lambda}_{11} \sim \lambda_{11} 
\Lambda /M$. 

I assume that $S$, $A^{\prime \prime}$, and $A$ acquire the vevs 
\begin{equation}
S= s (1,1,1,-\frac{3}{2},-\frac{3}{2}) \otimes \left( 
\begin{array}{cc}
1 & 0 \\
0 & 1 
\end{array} \right) ,
\hbox{ }
A^{\prime \prime}= 
(a^{\prime \prime} ,a^{\prime \prime},a^{\prime \prime},
b^{\prime \prime},b^{\prime \prime}) \otimes \left(
\begin{array}{cc}
0 & -1 \\
1 & 0
\end{array} \right), 
\end{equation}
\begin{equation}
A=
(a,a,a,b,b) \otimes \left(
\begin{array}{cc}
0 & -1 \\
1 & 0
\end{array} \right).
\end{equation}
These vevs break 
$SO(10) \rightarrow SU(3) \times SU(2) \times U(1)_Y \times U(1)_X$.
The spinor field ${\bf 16}$ is assumed to acquire a 
vev $\chi$ in the $SU(5)$-singlet 
direction\footnote{The $D-$flatness condition for 
$SO(10)$ requires
the vevs of ${\bf 16}$ and $\overline{{\bf 16}}$ to be equal.}.
The unbroken gauge group is then $SU(3) \times SU(2) \times U(1)_Y$.
It is argued below that
the superpotential guarantees that the vevs of $A$, $\sigma$,
$A^{\prime \prime}$, $S$, and ${\bf 16}$ are
naturally on
the order of $\Lambda \sim M_{GUT}$.

The doublets and triplets in $T_1$ are split using the DW mechanism \cite{dw}.
The $F_A$ equations 
$(\tilde{\lambda}_9 s + \tilde{\lambda}_{10} \sigma)a=0$ and 
$(-\frac{3}{2} \tilde{\lambda}_9 s + \tilde{\lambda}_{10} \sigma)b=0$
with $s\neq 0$ forces either $a$ or $b$ to vanish; it is a 
discrete choice. The DW 
mechanism for giving the triplets in the ${\bf 5}_{1,2}$ and 
$\overline{{\bf 5}}_{1,2}$ Higgs fields GUT-sized masses 
requires that $b=0$. I assume that this 
minimum was selected in the early history of the universe. With this 
choice, the mass matrix for the coloured triplets in the ${\bf 5}_{1,2}$ and 
$\overline{{\bf 5}}_{1,2}$ Higgs fields is 
\begin{equation} {\cal M}=\left(
\begin{array}{cc}
0 & -i \lambda_1 a \\
i \lambda_1 a & \tilde{\lambda}_2 \sigma + \tilde{\lambda}_3 s 
\end{array} \right)
\end{equation}
in the $(T_1, T_2)$ basis. Since the diagonal element is suppressed
by a factor of $O( \Lambda/ M)$ relative to the off-diagonal element, 
the coloured triplets form two Dirac particles with masses 
$m_T \sim \lambda _1 a \sim \lambda_1 \Lambda \sim \Lambda$. 
The mass matrix for the 
4 electroweak doublets in $T_1$ and $T_2$ only has an entry for 
$T_2( {\bf 2}) T_2 ({\bf \bar{2}})$ since $b=0$. The mass of the Dirac heavy 
doublet is $\tilde{\lambda}_2 \sigma -3 \tilde{\lambda}_3 s/2
\sim \Lambda^2 /M$. The two electroweak doublets in $T_1$ and 
$T_2$ are massless, and are identified as the Higgs fields responsible for 
giving mass to the up-type and down-type quarks of the SM.

I note that the magnitude of the elements of ${\cal M}$ 
has a structure that is favourable for the suppression of the proton 
decay rate. In particular, the diagonal element is suppressed by a factor 
of $O( \Lambda/ M)$ relative to the off-diagonal element, reflecting 
the fact that the diagonal entry arises from a non-renormalizable 
operator in the fundamental theory. If the SM 
fermions only couple to $T_1$, then  
the proton decay amplitude from the exchange of the heavy 
coloured Higgsinos is proportional to ${\cal M}^{-1}_{11}$.
In this case the matrix element is 
$(\tilde{\lambda}_2 \sigma + \tilde{\lambda}_3 s)/(\lambda_1 a)^2
\sim \Lambda^2 /M$. This  
results in a decay rate that is approximately 
$(\Lambda / M)^2 \sim 10^{-3}$ 
times the unsuppressed rate. 

This is suffucient to suppress the dangerous Higgsino-exchange proton 
decay operator to a level that may be observable at SuperKamiokande.  
To obtain the four-fermion operator responsible for the nucleon decay, 
the operator gotten by integrating 
out the coloured triplet Higgsinos must be dressed with a vertex function 
involving either internal wino 
or gluino propagators. 
As emphasized in Reference \cite{babu1}, the gluino-dressed amplitude is
comparable to the wino-dressed amplitude if $v_u /v_d \equiv \tan \beta$ 
is large. 
Since  
$\tan \beta \sim m_t /m_b \sim 40$ is naturally predicted within an 
$SO(10)$ GUT, the decay mode $\hbox{p} \rightarrow K^{0} \mu ^{+}$ 
may be 
competitive with the (wino-dressed) neutrino decay modes \cite{babu1}.

The dominant decay modes for the wino-dressed operator
are $\hbox{p} \rightarrow K^{+} \bar{\nu}_u$ and 
$\hbox{n} \rightarrow K^{0} \bar{\nu}_u$ \cite{hitoshi}.
To obtain an estimate for 
the nucleon lifetime in this model, I rescale their result  
for the 
lifetime of the nucleon by a factor of $(M/ \Lambda)^2 $. 
The result is 
\begin{equation}
\tau(\hbox{n} \rightarrow K^{0} \bar{\nu}_u) \sim  
1.0 \times 10^{32} \times \left( \frac{M}{30 \Lambda}  
 \frac{0.0058 \hbox{GeV}^3}{\beta} 
\frac{M_{H_c}} {10 ^{16} \hbox{GeV} } 
 \frac{ \hbox{TeV}^{-1} } {f(\tilde{u},\tilde{d})+
f(\tilde{u},\tilde{e})} \right)^2 \hbox{yrs}.
\label{nucleon}
\end{equation}
The function $f$ is obtained by dressing the external squarks with 
wino propagators to obtain a four-fermion operator. 
It is computed in Reference \cite{nath}, and 
depends on the sparticle spectrum. In the limit that the squark 
mass, $m_{\tilde{Q}}$, and slepton mass, $m_{\tilde{L}}$, are much larger 
than the wino mass, $m_{\tilde{w}}$, $f \sim m_{\tilde{w}} / m^2_{\tilde{X}}$,
with $m_{\tilde{X}}$ the larger of $m_{\tilde{Q}}$ and $m_{\tilde{L}}$. 
The hadronic matrix element $\beta$ is  
defined in Reference \cite{hitoshi}, and $M_{H_c}$ is the mass of the 
coloured triplets.  
Requiring that $M$ not exceed the Landau pole of the $SO(10)_{GUT}$
group 
implies that $M/ \Lambda \ltap \hbox{ }30-70$.
This requirement of consistency also strongly constrains the presence 
of any additional matter content (this is also discussed below). 
This suggests that the Yukawa 
couplings of the SM fermions
to the Higgs doublets
are generated close to the GUT scale, a crucial assumption 
required to obtained the limit quoted in Equation \ref{nucleon}.
To obtain realistic quark and lepton masses in an $SO(10)$ model though, 
these Yukawa couplings probably arise from higher-dimensional 
operators \cite{hall5}. In this case the flavour structure of 
the coloured-triplet Higgs to matter may differ from 
the electroweak doublet 
couplings to matter, 
thereby altering the predicted 
lifetime \cite{babu1}. For this reason, the 
result quoted in Equation \ref{nucleon}
should be treated as an estimate. This estimate is to be compared
with the existing
experimental limit of
$\tau(\hbox{n} \rightarrow K^{0} \bar{\nu}_u)> .86 \times 10^{32}$
years \cite{pdg}.
So the nucleon lifetime is naturally 
suppressed to a phenomenologically interesting
level. 

Next I discuss the expected size of the vevs and the mass 
spectrum. The $F-$flatness equations are (setting $b=0$) 
\begin{equation}
\det( S+ A^{\prime \prime} + \sigma / \sqrt{10} ) - B \overline{B}= 
\Lambda ^{10} ,
\end{equation}
\begin{equation}
{\cal A} B=0 \hbox{ }, \hbox{ } {\cal A} \overline{B}=0,
\end{equation}
\begin{equation}
0=F_A = (\tilde{\lambda}_9 s + \tilde{\lambda}_{10} \sigma) a,
\end{equation}
\begin{equation}
0=F_{16}= \left( \tilde{\lambda}_{16} \sigma + \tilde{\lambda}_4 
(3 a^{\prime \prime} + 2 b ^{\prime \prime})\right) \chi,
\end{equation}
\begin{equation}
0=F_{\sigma}= \tilde{\lambda}_{16} \chi ^2 -3 \tilde{\lambda}_{10} a^2
+\tilde{\lambda}_5 \Lambda \sigma + \frac{2}{\sqrt{10}} {\cal A} K
\left( 3 u +2 v \right), 
\end{equation}
\begin{equation}
0=F_{{A^{\prime \prime}}_{3}}= \tilde{\lambda}_4 \chi ^2 -
2 \tilde{\lambda}_7 \Lambda a^{\prime \prime} +2 {\cal A} K A, 
\end{equation}
\begin{equation}
0=F_{{A^{\prime \prime}}_{2}}= \tilde{\lambda}_4 \chi ^2 -
2 \tilde{\lambda}_7 \Lambda b^{\prime \prime} +2 {\cal A} K B, 
\end{equation}
\begin{equation}
0=F_S=\tilde{\lambda}_6 \Lambda s - 
\frac{2}{5} \left( \frac{1}{2} \tilde{\lambda}_9 a^2
- {\cal A} K (u-v) \right),
\end{equation}
where $K \equiv \det (S+ A^{\prime \prime}+ \sigma/ \sqrt{10})
=(u^2+A^2)^{-3} (v^2+B^2)^{-2}$.
The functions $u$, $v$, $A$ and $B$ are  
\begin{equation}
u=\frac{ \sigma / \sqrt{10} + s}{(\sigma / \sqrt{10} + s)^2 +
{a^{\prime \prime}}^2 } \hbox{ } , \hbox{ } 
v=\frac{\sigma / \sqrt{10} - 3 s /2}
{(\sigma / \sqrt{10} - 3 s/2)^2 + {b^{\prime \prime}}^2},
\end{equation}
\begin{equation}
A= \frac{ a^{\prime \prime}}
{(\sigma / \sqrt{10} + s)^2 +{a^{\prime \prime}}^2} \hbox{ } , \hbox{ }
B= \frac{ b^{\prime \prime}}
{(\sigma / \sqrt{10} -3 s/2)^2 +{b^{\prime \prime}}^2}.
\end{equation}
An inspection of these 
equations 
also indicates that without the operators $\Lambda S^2$, $\Lambda \sigma^2$ and
$\Lambda A^{\prime \prime} A^{\prime \prime}$, the $F-$flatness 
equations would only constrain the values of  
${\cal A}$, $\chi^2$ and $a^2$ in the combination  
$\chi^2 /{\cal A}$ and $a^2/ {\cal A}$. Thus one of these vevs  
would be
unconstrained. As a result, not all the particle masses would be 
fixed by the input parameters. 
This problem is avoided by 
including the $(Q \overline{Q})^2$ operator 
in the fundamental theory. In this case, a new solution cannot be 
gotten by rescaling ${\cal A}$, 
with the $\tilde{\lambda}_i$ and $\Lambda$ fixed,  
and rescaling the vevs of any of the fields, thus indicating that 
$a^2$, $\chi ^2$ and ${\cal A}$ are fixed by the input parameters. 

I now argue that these equations 
fix the vevs of $S$, $A$, $\sigma$ and $A^{\prime \prime}$
to be on the order of $\Lambda$, without any fine tuning of the couplings 
in the fundamental theory. 
By redefining ${\cal A} =(\Lambda /M) \tilde{{\cal A}}$ the 
$F_i=0$ equations now contain
an overall factor of $\Lambda / M $ if the expected relation between
the superpotential couplings in the fundamental and confined theories
is valid. As a result the $F_i$ equations
no longer contain any small dimensionless couplings.
The expected
solution to
this new set of equations is then $\chi$, $\sigma$, $a^{\prime \prime}$,
$b^{\prime \prime}$, $a$ $\sim s$ and
$\tilde{{\cal A}} \sim \Lambda^{-7}$. The confinement equation fixes 
$s \sim \Lambda$.
Therefore all the vevs are $v \sim \Lambda$ and
${\cal A} \sim (\Lambda / M) \Lambda^{-7}$.
This result is not 
obvious ${\it a \hbox{ } priori}$, since the superpotential couplings
appearing in the $F$ equations are
suppressed by powers of $\Lambda/ M$. 
A slightly more rigorous argument,
also showing that ${\cal A} \neq 0$, is presented in Appendix B. 
Two numerical solutions which supports these arguments are also given in 
Appendix B.
These expectations for the size of the couplings, ${\cal A}$, and 
vevs will be important below in estimating the mass spectrum.

The superpotential for this model contains
enough operators to give superheavy masses
to all the particles that should be heavy. 
The results of computing the mass matrices assuming a canonical 
Kahler potential are given in Appendix B, and are summarized here. 
The particles have masses at one 
of three scales: $m_L \equiv \Lambda^4/ M^3$; $m_I \equiv \Lambda^2 /M$; and 
$\Lambda$. The naive expectation is that all the particles have a mass 
$m \sim m_I$. This is because all the vevs are $O(\Lambda)$, and 
the mass matrices are linear in
the superpotential couplings which contain a factor $\Lambda/M$, and in
the parameter ${\cal A}$ which also contains a factor of $\Lambda/M$. 

This expectation turns out to be correct
except for a $u_L \sim ({\bf \overline{3}},{\bf 1},-2/3)$
and $\overline{u}_L \sim u^{\dagger} _L$, which acquire a Dirac mass
from the superpotential operator 
$(A^{\prime \prime} A)_{ij} \overline{ {\bf 16}}\Sigma_{ij} {\bf 16}$.
These fields are massless in the absence of this operator 
for the following reason. 
The $SU(5)$ decomposition 
of $A={\bf 24}+{\bf 10} +\overline{{\bf 10}}+{\bf 1}$. 
This clearly contains a $u \hbox{ }\epsilon \hbox{ }{\bf 10}$ and 
$\overline{u}  \hbox{ }\epsilon \hbox{ } 
\overline{{\bf 10}}$.  
The only possible source for a mass term for these 
fields is given by $W_{DW}$.
Further, since $S$ does not contain a $u$ and $\overline{u}$, this mass 
term must occur from setting $S$ and $\sigma$ to their vevs. The resulting 
mass is proportional to $\tilde{\lambda}_9 s + \tilde{\lambda}_{10} \sigma$.
The $DW$ form for $A$ and $F_A=0$, however, forces 
this quantity to vanish \footnote{The same argument also implies that the 
Majorana mass term for the ${\bf 8}$ in $A$ vanishes. These fields, however, 
acquire a Dirac mass with the ${\bf 8}$ $\epsilon$ $S$.}. 
The addition of the operator $\hbox{tr} A^4/ M$ does not change the 
conclusion of this argument. 
The mass of these fields 
is gotten therefore from the $M^{-2}$ suppressed operator.
The result of  
a computation of the mass spectrum, presented in Appendix B,
 implies that the naive
expectation for their
mass is $m \sim m_L$. 

The particle content of the fields with 
mass $m \sim m_I$ is now enumerated. The complete $SU(5)$ representations at 
this scale are: $1 \times  ({\bf 10} 
+\overline{{\bf 5}})+
1 \times (\overline{{\bf 10}} 
+ {\bf 5}) +  2 \times {\bf 24}+ 1 \times ({\bf 15}
+\overline{{\bf 15}})$.
At the scale $m_I$ there is also a split ${\bf 24}$, with 
SM quantum numbers ${\bf 8} \equiv ({\bf 8},{\bf 1},0)$ and 
${\bf 3}\equiv ({\bf 1},{\bf 3},0)$. There are also some leftover 
fields, which together with $u_L$ and $\overline{u}_L$, form a
complete ${\bf 10}+{\bf \overline{10}}$ of $SU(5)$. These leftover 
fields have a mass $m \sim m_I$.
The representations in the $SO(10)$ ${\bf 10_1}+{\bf 10_2}$ are split by the 
DW mechanism. One pair of electroweak doublets is massless and  
are the Higgs fields responsible for giving mass to the up-type quarks, 
down-type quarks, and leptons. 
The other doublet fields, 
$h \equiv ({\bf 1},{\bf 2},-1/2)$ and 
$\overline{h}\equiv ({\bf 1},{\bf 2},1/2)$, 
acquire a Dirac mass $m_h \sim m_I$. 
There are also a number of gauge singlets
which acquire masses $m \sim m_I$.

The triplets in the $SO(10)$ 
${\bf 10_1}+{\bf 10_2}$, 
$2 \times (\overline{{\bf 3}},{\bf 1},1/3)+2 \times ({\bf 3},{\bf 1},-1/3)$,  
acquire masses $O (\Lambda)$. The 33 Nambu-Goldstone bosons multiplets 
acquire a mass $m \sim \Lambda$ from the super-Higgs mechanism.

The incomplete $SU(5)$ representations 
affect the prediction for $\sin ^2 \theta_W$, which I now discuss.
I find using the usual one-loop computation 
that the light particles shift the prediction for $\sin ^2 \theta_W$
by an amount 
\begin{equation}
\Delta \sin^2\theta_W =-\frac{\alpha_{em}}{2 \pi} \left(
\ln \frac{m_I}{m_L} - \frac{4}{5} \ln \frac{\Lambda}{m_I} \right).
\end{equation}
The first term is the contribution from $u_L$ and 
$\overline{u}_L$; these fields only contribute between $m_L$ and 
$m_I$, since above the mass scale $m_I$ they fit into a
complete ${\bf 10}+{\bf \overline{10}}$ of $SU(5)$. The second 
term is the sum of the contributions from ${\bf 8}$, ${\bf 3}$, 
$h$ and $\overline{h}$.
As is evident, for $m_L < m_I$ 
there is an $O(1)$ cancellation 
between the two contributions. Since 
$m_L$ arises from a higher dimensional operator than does $m_I$, 
$m_L < m_I$ applies for this model. It is then reasonable to 
expect that the $O(1)$ cancellation occurs.
Inserting the naive 
expectation $m_L \sim \Lambda^4/M^3$ and $m_I \sim \Lambda^2/M$,
gives
\begin{equation}
\Delta \sin^2\theta_W \sim-5 \times 10^{-3} \times 
\frac{ \ln M / \Lambda} {\ln 30}.
\label{sosin}
\end{equation}
As is shown below, requiring that the $SO(10)_{GUT}$ not have a Landau 
pole below $M$ restricts $M/ \Lambda \ltap \hbox{ }30-70$. With 
this constraint, 
the shift in $\sin^2 \theta_W$ is consistent with the measured value, once 
other theoretical uncertainties are considered. The largest 
of these are uncalculable 
threshold corrections
from the light (approximately) complete $SU(5)$ representations. 
Since the contribution of each multiplet 
is naively $\alpha _{em}/ 2\pi \times O(1)$, the large size of 
the light representations could result in a correction that is 
comparable or larger than the correction given in Equation \ref{sosin}. 
 
I now argue that any 
``gravitational smearing'' \cite{hall2} of the couplings at the GUT scale 
is small in
this model.
First, the only possible dimension$-$4 operator in the superpotential
involving the $SO(10)_{GUT}$ chiral gauge multiplet $W_{ij}$ 
is $A_{ij} W_{jk} W_{ki} /M$. This, however, vanishes due to the
anti-symmetry
of $A$. Next, the operators $ g_S S W W /M$ and $ g_{\sigma} \sigma W W /M$
are allowed. The vev of $\sigma$ does not break $SU(5)$, so it
only results in a common shift of the gauge couplings.
The shift is tiny since $g_{\sigma} \sim \Lambda /M$.
The vev of $S$ does break $SU(5)$, so this operator results
in a tree-level correction to the unification of the couplings.
An estimate for the shift in $\sin ^2 \theta_W$ that this incurs 
is 
\begin{equation}
\Delta \sin^2 \theta_W \sim \pm 10^{-3} g_S \frac{30 s}{M}.
\end{equation} 
It is expected that $g_S \sim \Lambda/ M$ since 
this operator occurs from a dimension$-$4 operator in the superpotential
of the fundamental theory.
So this results in a tiny shift to $\sin^2 \theta_W$.
Finally, operators only involving ${\bf 16},
{\overline{\bf 16}}$ and $WW$ 
are also suppressed by an extra factor of $\Lambda /M$.
The vev of ${\bf 16}$ does not break $SU(5)$,
so this operator only results in a tiny common shift to the gauge couplings.

An upper limit to $M$ is given by 
the value of the Landau pole of the $SO(10)$ GUT
gauge coupling.  This model is not asymptotically-free 
above the GUT scale since it contains a large particle content.
More problematic though, is the fact 
that most of the particle masses are a factor of $\Lambda /M$ below 
the GUT scale. 
While this particle content does
not result in a large shift to 
$\sin ^2 \theta_W$ since they mostly form complete $SU(5)$ representations, 
the matter content does increase the value of 
$\alpha_{GUT}$.
The value of $\alpha_{S0(10)}(M)$, using naive one-loop running and 
with tree-level matching, and including the contribution of 3 ${\bf 16}$s
of the SM, 
is 
\begin{equation}
\alpha^{-1} _{S0(10)}(M)=24
-\frac{3}{22 \pi} \left((2+\frac{5}{3}) \ln \frac{\Lambda}{m_L}
+(93-\frac{5}{3}) \ln \frac{\Lambda}{m_I}\right)-\frac{16}{2 \pi} 
\ln \frac{M}{\Lambda}.
\label{alphaM}
\end{equation}
The second term is the contribution from $u_L+\overline{u}_L$,
the third term is the 
contribution from the particles with mass $m_I$, and the last term 
is the contribution from the $SO(10)$ particle content above $\Lambda$.
Inserting $m_L \sim \Lambda^4/ M^3$ and $m_I \sim \Lambda^2 /M$, 
the limit is 
\begin{equation}
\frac{M}{\Lambda} \ltap \hbox{ } 31.
\end{equation}
This implies $M \sim .6-1 \times 10^{18}$ GeV.  
I note, however, that this limit is sensitive to the actual spectrum. 
For example,  
if the naive expectation underestimates the spectrum by a factor of 4, then the 
limit increases to $M / \Lambda \ltap \hbox{ } 75$. This corresponds to 
$M \sim 1-2 \times 10^{18}$ GeV. 

The Landau pole limit also 
strongly constrains any modifications to the model.
For example, adding to the model either an extra adjoint $A^{\prime}$ which
acquires a mass at $2 \times M_{GUT}$, or an extra 
${\bf 16}^{\prime}+\overline{\bf 16}^{\prime}
+{\bf 10}^{\prime}+
{\bf 10}^{\prime \prime}$ which all acquire a mass $M_{GUT}$ 
restricts $M/ \Lambda \ltap 20$.
The presence of $N_5$ additional $SU(5)$ ${\bf 5}+\overline{\bf 5}$
multiplets is also strongly constrained by this requirement of consistency.
These fields would be required, 
for example, in any low-energy physics 
that is responsible for 
the origin of supersymmetry or flavour symmetry breaking.
Requiring $M/ \Lambda > 20$ implies that 
the mass $M_5$ of these multiplets satisfies 
\begin{equation}
 N_{{\bf 5}+\overline{\bf 5}} \ln M/ M_5 \ltap \hbox{ }18.
\label{c1}
\end{equation}
In particular: $N_{{\bf 5}+\overline{\bf 5}}=1$ is marginally 
allowed if $M_5= 10^{10}$ GeV; $N_{{\bf 5}+\overline{\bf 5}}=2$ is 
marginally allowed if $M_5= 10^{14}$ GeV.
These constraints are weakened if the naive estimate, $\Lambda^2 /M$, 
for the chiral 
GUT spectrum underestimates the spectrum by a factor of 4. In this case,
\begin{equation}
 N_{{\bf 5}+\overline{\bf 5}} \ln M/ M_5 \ltap \hbox{ }45,
\label{c2}
\end{equation}
for $M/ \Lambda >20$. In particular: $N_{{\bf 5}+\overline{\bf 5}}=2$ is 
allowed for $M_5= 10^{10}$ GeV; $N_{{\bf 5}+\overline{\bf 5}}<5$ is 
required for $M_5= 10^{14}$ GeV. 
Either direct or indirect evidence for additional chiral content that 
does not satisfy Equations \ref{c1} or Equations \ref{c2} would 
strongly disfavour this model. 

I conclude this Section with a few comments 
about the consistency of neglecting certain 
operators in the superpotential.
The superpotential terms $\sigma A_{ij}
\overline{{\bf 16}} \Sigma_{ij} {\bf 16}$ or $S_{ik} A_{kj}
\overline{{\bf 16}} \Sigma_{ij} {\bf 16}$ must be absent to
avoid ruining the DW form for $A$.
These operators would contribute to
$F_A(2)$, forcing a non-vanishing value for $b$.
These operators are present in
the low-energy theory if the operators $\hbox{tr}( \overline{Q} Q )
A \overline{{\bf 16}} \Sigma {\bf 16}$ or $( \overline{Q} Q )_S A
\overline{{\bf 16}} \Sigma {\bf 16}$ are present in the
superpotential of the fundamental theory.
Any symmetry which forbids these dangerous operators also forbids the
operator $(A^{\prime \prime} A)_{ij} \overline{ {\bf 16}}\Sigma_{ij} {\bf 16}$.
This option is not viable since this operator is
required to give mass to a
$(\overline{{\bf 3}}, {\bf 1},-2/3)+$h.c. fields.
(The DW form for $A$, however, is unaffected by the presence of this operator
since it does not contribute to the $F_i$ equations.)
So I must assume that the dangerous operators are
not present in the fundamental
theory. 
The perturbative non-renormalisation theorems then 
guarantee that these operators 
will not be generated, at least in perturbation theory. 
This argument does not exclude the possibility that these dangerous 
operators could be generated by the  
non-perturbative dynamics of the $SU(10)_S$ or $SO(10)_{GUT}$ 
groups.     
By combining the requirement of holomorphy of the superpotential 
with some anomalous fake $U(1)$ symmetries
it is possible to exactly show, however, 
that if these operators are initially absent in the
high-energy theory they will not be generated as the cutoff is lowered.
In particular, it can be shown that the coefficient of a dangerous
operator at a lower cutoff is only proportional to its initial value; i.e.
it is independent of $\Lambda_{SU(10)} /M$, $\Lambda_{SO(10)}/M$ and
all the other superpotential couplings. I then see no reason for these
dangerous operators to be generated by the confining dynamics.

\section{Acknowledgements}
The author would like to thank C. Cs\'aki, T. Moroi and J. Terning for 
valuable discussions, and 
I. Hinchliffe and M. Suzuki for a careful reading of the manuscript.
This work was supported in part
by the Director, Office of Energy
Research, Office of High Energy and Nuclear Physics,
Division of High
Energy Physics of the U.S. Department of Energy
under Contract
DE-AC03-76SF00098.
The author also thanks the Natural Sciences 
and Engineering Research Council of Canada for their support. 

\section {Appendix A: $SU(6) \times SU(6)_{GUT}$}

First I discuss the existence of a solution to the $F_i=0$ 
equations with all vevs of $O(\Lambda)$ and 
${\cal A} \sim (\Lambda / M) \Lambda ^{-3}$. The second part 
of this Appendix contains the results of calculating the 
mass spectrum, assuming a canonical Kahler potential. 
 
Since the $F_i=0$ equations are linear in $v_H ^2 $ and 
$v^2 _N$, it is straightforward to solve for them in terms 
of $\sigma$ and $v_{\Sigma}$. The remaining two equations 
determine ${\cal A}\neq 0$ and $x \equiv \sigma /v_{\Sigma}$. 
In particular, $x$ is the solution to 
\begin{equation}
\beta x^2 - (\sqrt{6} \beta - \alpha - \gamma) x -\sqrt{6} \alpha 
+12 \beta =0,
\label{x}
\end{equation}
where $\alpha \equiv - \tilde{\lambda}_1$, 
$\beta \equiv - \tilde{\lambda}_3 \tilde{\lambda}_4 M / (12 g \Lambda)$, 
and $\gamma \equiv -\tilde{\lambda}_2 
-24 \tilde{\lambda}_5 \beta / \tilde{\lambda}_4$. 
Since $ \beta \sim \alpha \sim \gamma \sim \Lambda /M$, Equation 
\ref{x} implies that   
$ \sigma \sim v_{\Sigma}$ is expected. The quantum constraint 
then fixes $v_{\Sigma} \sim \Lambda$. It follows from $F_H=0$ that 
$v^2 _N =- \tilde{\lambda}_3 M \sigma /(12 g)$ is $O(\Lambda^2)$. 
Next, $v^2 _H =-(M v_{\Sigma} /g) (\tilde{\lambda}_5 x -\tilde{\lambda}_4)$ 
is also $O(\Lambda ^2)$. Finally, either $F_{\Sigma}=0$ or 
$F_{\sigma}=0$ determines ${\cal A} \sim (\Lambda /M ) \Lambda^{-3}$.
 
The non-Nambu-Goldstone multiplet fields charged under the SM, 
with the exception of the SM Higgs doublets, are all 
contained in $\Sigma$ and $\Sigma_N$. Since these fields acquire 
their mass from the $SU(4) \times SU(2)$ preserving 
vevs of $\Sigma$, $\Sigma_N$ or 
$(H \overline{H})$, it is conveinent to classify the mass 
spectrum according to the $SU(4) \times SU(2)$, rather 
than the $SU(3) \times SU(2) \times U(1)_Y$, charge assignments. 

The mass matrix for the $Q\sim ({\bf 4},{\bf 2})$ and 
$\overline{Q} \sim (\overline{{\bf 4}},{\bf 2})$ fields  
(after some algebra using the $F_i=0$ equations)
in the $(\Sigma, \Sigma_N)$ basis is
\begin{equation}
M_{\overline{Q} Q} =\left( \begin{array}{cc} 
-{\cal A} K ab +\tilde{\lambda}_1 \Lambda & - \tilde{\lambda}_4 v_N \\
- \tilde{\lambda}_4 v_N & \tilde{\lambda}_4 v_{\Sigma} 
\end{array} \right).
\end{equation}
By using the $F_i=0$ equations it 
can be verified that this matrix annihilates the state $(v_{\Sigma}, v_N)$,
which is a Nambu-Goldstone
boson of the gauge symmetry breaking.
The massive eigenvalue is non-zero and naively $m_Q \sim \Lambda^2 /M$.

The mass matrix for the $({\bf 15},{\bf 1})$ fields 
(after some algebra using the $F_i=0$ equations) 
in the $(\Sigma, \Sigma_N)$ basis is
\begin{equation}
M_{{\bf 15} }= \left( \begin{array}{cc} 
- {\cal A} K a^2 +\tilde{\lambda}_1 \Lambda & 2 \tilde{\lambda}_4 v_N \\
2 \tilde{\lambda}_4 v_N & 4 \tilde{\lambda}_4 v_{\Sigma} 
\end{array} \right).
\end{equation}
It can be shown after some algebra 
that the determinant of this matrix is 
$-4 \tilde{\lambda}_4 {\cal A} K a v_{\Sigma} (a-b)$
. This is non-zero since $v_{\Sigma}  \neq 0$ 
implies that $a \neq b$.   
The expected masses for the two eigenvalues is then 
$m_{{\bf 15}} \sim \Lambda^2 /M$.

The mass matrix for the $({\bf 1}, {\bf 3})$ fields  
(after some algebra using the $F_i=0$ equations)
in the $(\Sigma, \Sigma_N)$ basis is
\begin{equation}
M_{{\bf 3}}=\left( \begin{array}{cc}
- {\cal A} K b^2 +\tilde{\lambda}_1 \Lambda & -4 \tilde{\lambda}_4 v_N \\
-4 \tilde{\lambda}_4 v_N & -2 \tilde{\lambda}_4 v_N \\
\end{array} \right).
\end{equation}
It can be shown that the determinant of this matrix is
$-\sqrt{6} \tilde{\lambda}_4 \Lambda b v ^2 _{\Sigma} 
(3 \beta x^2- 5 \beta \sqrt{6} x +\sqrt{6} \alpha)
$. A comparison of this result with Equation \ref{x} 
indicates that it is non-vanishing for generic values of the 
$\tilde{\lambda}_i$s. 
The expected masses for the two eigenvalues is then
$m_{{\bf 3}} \sim \Lambda^2 /M$.

\section {Appendix B: $SU(10) \times SO(10)_{GUT}$}

\underline{{\it Arguing that all the vevs 
are of order $\Lambda$}} ; \underline{{\it Numerical solution}}

In this case I am only concerned about whether a discrete solution with
all ${\cal A}$, $a$, $a^{\prime \prime}$, $b^{\prime \prime}$, $\sigma$, 
$s$ and 
$\chi$ non-zero exists. 
This result
is obtained by showing
that if $s \neq 0$, then ${\cal A} \neq 0$ and all other vevs are
comparable to $s$. Then the non-vanishing of ${\cal A}$ implies that
$B= \overline{B} =0$. The confinement condition then fixes $s \sim \Lambda$.
To begin, first note that $F_{A}$ fixes $\sigma \sim s$. The $F_{16}$ equation
implies that $3 a^{\prime \prime} + 2 b^{\prime \prime} \sim \sigma \sim s$.
Thus either $a^{\prime \prime} \sim b^{\prime \prime} \sim s$, or
$b^{\prime \prime} \ll  a^{\prime \prime} \sim s
$ $($or $a^{\prime \prime} \ll b^{\prime \prime} \sim s)$. I next argue that
the last two cases do not occur. In the first case,
$b^{\prime \prime} \ll  a^{\prime \prime}$, so that $B \ll A$.
Next, the two $F_{A^{\prime \prime}}$
equations are inconsistent if either
${\cal A}K A \ll \tilde{\lambda}_7 a^{\prime \prime}$ or
${\cal A}K A \gg \tilde{\lambda}_7 a^{\prime \prime}$.
So ${\cal A}K A \sim \tilde{\lambda}_7 a^{\prime \prime}$ and
$\tilde{\lambda}_7 b^{\prime \prime} \sim
\chi^2 \ll \tilde{\lambda}_7 a^{\prime \prime}$ is the only consistent
solution to the two $F_{A^{\prime \prime}}$ equations.
Thus if $b^{\prime \prime} \ll a^{\prime \prime}$, $F_{16}$
fixes $a^{\prime \prime} \sim s$ up to small corrections of
$O(b^{\prime \prime})$. Similarly, the first $F_{A^{\prime \prime}}$ fixes
${\cal A}$ up to small corrections. But now the two equations $F_{\sigma}$
and $F_S$ each determine $a \sim s$; these two equations for $a$ cannot
in general be simultaneously satisfied. Therefore, $b^{\prime \prime} \ll
a^{\prime \prime}$ is not a viable (supersymmetric) solution. The
argument against $a^{\prime \prime} \ll b^{\prime \prime}$ is similar.
Therefore $a^{\prime \prime} \sim b^{\prime \prime}$. Next suppose that
${\cal A}= 0$. Then $F_{A^{\prime \prime}}$ fixes
$a^{\prime \prime}=b^{\prime \prime}$, and together with $F_{16}$ and
$F_A$, determines $\chi \sim s$. But now there are two remaining equations,
$F_S$ and $F_{\sigma}$, for one unknown, $a$. More concretely,
$a^2 =5 (\tilde{\lambda}_6 / \tilde{\lambda}_9) \Lambda s$ and
$a^2= (\tilde{\lambda}_5-\frac{2}{5} \tilde{\lambda}_7
{\tilde{\lambda}}^2 _{16}/
{\tilde{\lambda}}^2 _{4}) \Lambda \sigma /3 \tilde{\lambda}_{10}$.
In general, these two equations will not be satisfied; therefore
${\cal A} \neq 0$. The vev $a$ can be eliminated from $F_{S}$ and
$F_{\sigma}$; the remaining equation, together with
$F_{A^{\prime \prime}}$ and $F_{16}$ may be used in principle
to determine $\chi, a^{\prime \prime}, b^{\prime \prime} \sim s$ and
also fix ${\cal A}$. ($\chi^2 \ll \Lambda a^{\prime \prime} $ is not
possible; $F_A$, $F_S$, $F_{A^{\prime \prime}}$, $F_{16}$ and $F_{\sigma}$ are
6 equations in only 5 unknowns: $\sigma$, $a^{\prime \prime}$,
$b^{\prime \prime}$,
$a$ and ${\cal A}$.) The $F_S$ equation will not in general be
satisfied with $a^2 \ll \Lambda s$ or $a^2 \gg \Lambda s$; since
${\cal A} K (u-v)$ is $O(\Lambda^2 s /M )$ and
 $\neq \tilde{\lambda}_6 \Lambda s$ in 
general, 
$F_S$ determines $a \sim s$. 

Two numerical solutions to the $F_i=0$ equations supports these arguments. 
In the first (I) solution, the input parameters are chosen to be : 
$\tilde{\lambda}_4 =0.01$, $\tilde{\lambda}_5 =0.02$, 
$\tilde{\lambda}_6 =0.03$, $\tilde{\lambda}_7 =0.04$, $\tilde{\lambda}_9 
=0.05$, $\tilde{\lambda}_{10}=0.06$ and $\tilde{\lambda}_{16}=0.045$.  
The solution,
in units of $\Lambda=1$, is
\begin{equation}
\sigma =-0.64, \hbox{ } s=0.77, \hbox{ } a^{\prime \prime} =0.50,
\hbox{ } b^{\prime \prime} = 0.70, \hbox{ } a=1.2, \hbox{ } \chi= 2.5,
\hbox{ } {\cal A} =-0.01.
\label{vevsol}
\end{equation}
In the second (II) solution, the input parameters are chosen to be : 
$\tilde{\lambda}_4 =0.0134$, $\tilde{\lambda}_5 =0.0123$,
$\tilde{\lambda}_6 =-0.03$, $\tilde{\lambda}_7 =0.0225$, $\tilde{\lambda}_9
=0.045$, $\tilde{\lambda}_{10}=0.0623$ and $\tilde{\lambda}_{16}=0.03657$.
The solution,
in units of $\Lambda=1$, is
\begin{equation}
\sigma =-0.62, \hbox{ } s=0.85, \hbox{ } a^{\prime \prime} =-0.14,
\hbox{ } b^{\prime \prime} = 1.1, \hbox{ } a=-0.87, \hbox{ } \chi=1.2,
\hbox{ } {\cal A} =0.04.
\end{equation}
These parameters are chosen to be small  
since $\tilde{\lambda} \sim \lambda \Lambda /M 
\sim 0.03 \lambda$ for $\Lambda /M \sim 1/30$. Aside from this feature, there 
is nothing special about this choice of superpotential couplings. 
As expected, all the vevs are $O(\Lambda)$ and ${\cal A} \sim (\Lambda /M) 
\Lambda ^{-7}$.   

\underline{{\it Detailed Mass Spectrum}}

The mass matrices presented here were computed assuming a canonical 
Kahler potential; this is suffucient to determine the rank of the matrix.

For future purposes it will be useful to note that 
the $F_{i}$ equations are invariant under the following
rescaling of couplings and fields:
\begin{equation}
(\tilde{\lambda}_4,\tilde{\lambda}_9,\tilde{\lambda}_{10},\tilde{\lambda}_{16})
\rightarrow ( g^{-2} \tilde{\lambda}_4, g^{-2} \tilde{\lambda}_9,
g^{-2} \tilde{\lambda}_{10}, g^{-2} \tilde{\lambda}_{16}),
\label{rs1}
\end{equation}
\begin{equation}
(\chi, a) \rightarrow (g \chi, g a) \hbox{ } ,
(a^{\prime \prime}, b^{\prime \prime}, s, \sigma, K) \rightarrow
(a^{\prime \prime}, b^{\prime \prime}, s, \sigma, K).
\label{rs2}
\end{equation}
Any coupling not listed is left invariant. This mapping relates
the solutions to the $F_i=0$ equations in two theories with 
different superpotential couplings which are related by this 
scale transformation.

The $u^{c} \sim({\bf \overline{3}},1,-2/3)$+h.c. mass matrix  
in the
$(A^{\prime \prime}, {\bf 16 (\overline{16})}, A)$ basis is, with
$\bar{\lambda} \equiv \tilde{\lambda}_{11} /M$,
\begin{equation}
M_{u^c, \overline{u}^c}=
\left( \begin{array}{ccc}
2 {\cal A} K (u^2+A^2)-2 \tilde{\lambda}_7 \Lambda &
2 i \tilde{\lambda}_4 \chi- 2 \bar{\lambda} a \chi &
i \bar{\lambda} \chi^2 \\
-2 i \tilde{\lambda}_4 \chi-2 \bar{\lambda} a \chi &
-4 \tilde{\lambda}_4 a^{\prime \prime} &
-2 \bar{\lambda} a^{\prime \prime} \chi \\
-i \bar{\lambda} \chi^2 & -2 \bar{\lambda} a^{\prime \prime} \chi & 0
\end{array} \right)
.
\end{equation}
Using the $F_i=0$ equations the reader
can verify that this matrix has only one zero eigenvalue.
The
product of the two non-zero
eigenvalues is given by the
coefficient of $O(e)$ in the
expansion of $\det (M_u -e{\bf 1})$.
This coefficient is
$\overline{\lambda}^2 \chi^2 (4 a^2+4 {a^{\prime \prime}}^2+ \chi^2)$.
Therefore, this matrix contains an extra massless particle in
the limit $\overline{\lambda} \rightarrow 0$.
With $\overline{\lambda} \neq 0$,
the naive expectation for this
product of eigenvalues is $(\Lambda /M)^4 \Lambda^2$.
The larger
eigenvalue is
$m_{u_H}=\tilde{\lambda}_4 ( 4 a^{\prime \prime} +\chi^2 / a^{\prime \prime})$,
and is approximately $\Lambda^2 /M$.  So the smaller eigenvalue is
$m_{u_L}=\bar{\lambda}^2 \chi^2 
(4 a^2+4 {a^{\prime \prime}}^2+ \chi^2)/m_{u_H}$.
The naive expectation for this quantity is $(\Lambda /M )^3 \Lambda$.

The mass matrix for $E^{c} \sim ({\bf 1}, {\bf 1},1)$ + h.c., in the
$(A^{\prime \prime}, {\bf 16(\overline{16})}, A)$ basis is
\begin{equation}
M_{E^c, \overline{E^c}}=
\left( \begin{array}{ccc}
2 {\cal A} K (v^2+B^2)- 2 \tilde{\lambda}_7 \Lambda &
2 i \tilde{\lambda}_4 \chi & i \bar{\lambda} \chi^2 \\
-2 i \tilde{\lambda}_4 \chi & -4 \tilde{\lambda}_4 b^{\prime \prime} &
-2 \bar{\lambda} b^{\prime \prime} \chi \\
-i \bar{\lambda} \chi^2 & -2 \bar{\lambda} b^{\prime \prime} \chi &
5 \tilde{\lambda}_9 s
\end{array} \right)
.
\end{equation}
Using the $F_i=0$ equations it can be verified that this mass matrix
has one zero eigenvalue.
The masses of
the other two states are $5 \tilde{\lambda}_9 s $ and
$- \tilde{\lambda}_4 (4 b^{\prime \prime}+ \chi^2/ b^{\prime \prime})$,
to lowest order in $\bar{\lambda} \Lambda$.

The mass matrix for the $Y \sim ({\bf 3}, {\bf 2}, -5/6)$ and
$X \sim (\bar{{\bf 3}}, {\bf 2}, 5/6)$  fields is given
in the $(A^{\prime \prime}, S, A)$ basis by
\begin{equation}
M_{YX}=
\left( \begin{array}{ccc}
-2 {\cal A} K (uv-AB)+2 \tilde{\lambda}_7 \Lambda & -2 i {\cal A} K(uB+vA) &
0 \\
-2 i {\cal A} K(uB+vA) & -2 {\cal A} K (uv-AB)+ 2 \tilde{\lambda}_6 \Lambda &
i \tilde{\lambda}_9 a \\
0 & i \tilde{\lambda}_9 a & -\frac{5}{2} \tilde{\lambda}_9 s
\end{array} \right).
\end{equation}
It can be verified, after some tedious algebra, that this matrix
has one zero eigenvalue. This matrix is therefore rank $2$. 
The masses of the other two states are $O(\Lambda^2 / M) $.

The $Q \sim ({\bf 3}, {\bf 2}, 1/6)$ and $\overline{Q}
\sim (\bar{{\bf 3}},{\bf 2},-1/6)$ mass matrix,
in the $(A^{\prime \prime}, S, {\bf 16(\overline{16})}, A)$ basis, is
\begin{equation}
M_{Q \overline{Q}}=\left( \begin{array}{cccc}
2 {\cal A} K (uv+AB)-2 \tilde{\lambda}_7 \Lambda & -2i {\cal A} K (Av-Bu) &
2i \tilde{\lambda}_4 \chi -\bar{\lambda} a \chi & i \bar{\lambda} \chi^2 \\
2 i {\cal A} K (Av-Bu) & -2 {\cal A} K (uv+AB)+2 \tilde{\lambda}_6 \Lambda &
0 & -i \tilde{\lambda}_9 a \\
-2 i \tilde{\lambda}_4 \chi -\bar{\lambda} a \chi & 0 &
-2 \tilde{\lambda}_4 (a^{\prime \prime}+b^{\prime \prime}) &
-\bar{\lambda} (a^{\prime \prime}+b^{\prime \prime}) \chi \\
-i \bar{\lambda} \chi^2 & i \tilde{\lambda}_9 a &
-\bar{\lambda} (a^{\prime \prime}+b^{\prime \prime}) \chi &
\frac{5}{2} \tilde{\lambda}_9 s
\end{array} \right).
\end{equation}
It can be verified that this matrix has at least one zero eigenvalue.
To verify that it has only one zero eigenvalue, it is suffucient to 
verify that the coefficient of $O(e)$ in the expansion of 
$\det(M_{Q \overline{Q}} -e {\bf 1})$ is non-vanishing. 
Since the entries proportional to $\bar{\lambda}$ 
result in a tiny perturbation to the
spectrum of $M_{Q \overline{Q}}$, it is suffucient to compute 
the $O(e)$ coefficient, call it $p$, while setting $\bar{\lambda}=0$.
In this case it is 
\begin{eqnarray}
p & = & 4 {\cal A} K \frac{(B u -A v)}{(u^2 +A^2) (v^2 +B^2)} 
(-2 \tilde{\lambda}_4 \tilde{\lambda}_6 ( u B^2 - (u-v) u v - A^2 v) \Lambda
\\ \nonumber
& & - \tilde{\lambda}_7 \tilde{\lambda}_9 (B (u^2+A^2)+A(v^2+B^2)) \Lambda
\\ \nonumber 
& & -\tilde{\lambda}_4 \tilde{\lambda}_9 ((A+B)^2+(u-v)^2)+{\cal A} K 
(\tilde{\lambda}_9 (A+B)-2 \tilde{\lambda}_4 (u-v)) (u^2+A^2)(v^2+B^2)).
\end{eqnarray}
If this vanishes at generic values for the couplings constants, then it must,
in particular,  
vanish for two solutions and sets of couplings constants that are 
related by Equations \ref{rs1} and 
\ref{rs2}. Under this scaling, however, $p \propto C \times (
c_1 g^{-2}+ c_2 g^{-4})$, with $C$, $c_1$ and $c_2$ functions of the 
initial vevs and couplings. 
This vanishes only if either $C=0$ or $c_1=0$ {\it and} 
$c_2=0$. The first condition implies $A v=B u$, and the second 
implies that $A+B=0$ {\it and} $u-v=0$. These conditions over-constrain 
the vevs, so they will  
not be satisfied at a generic solution. In particular, $p=(0.07)^3$ for 
the numerical solution (I)  
given by Equation \ref{vevsol}.  
The expected mass for the three massive eigenvalues is therefore 
$O(\Lambda^2 /M)$. 

The mass matrix for the coloured adjoints
$({\bf 8},{\bf 1},0)$ in the
$(A^{\prime \prime},S,A)$ basis is
\begin{equation}
M_{{\bf 8}{\bf 8}}=\left( \begin{array}{ccc}
- {\cal A} K(u^2-A^2)+\tilde{\lambda}_7 \Lambda &-2i {\cal A} K u A &
0 \\
-2i {\cal A} K u A & - {\cal A} K (u^2-A^2)+\tilde{\lambda}_6 \Lambda &
i \tilde{\lambda}_9 a \\
0 & i \tilde{\lambda}_9 a & 0
\end{array} \right)
\end{equation}
The determinant is $(\tilde{\lambda}_9 a)^2 ( \tilde{\lambda}_7 \Lambda 
-{\cal A} K (u^2 -A^2))$ and is non-vanishing.  
The size of the three masses is expected to be $m_8 \sim \Lambda^2/M$.
For the numerical
solution (I) in Equation \ref{vevsol}, this determinant is $(0.05)^3$.

The mass matrix for the $SU(2)$ adjoints
$({\bf 1},{\bf 3},0)$ in the
$(A^{\prime \prime},S,A)$ basis is
\begin{equation}
M_{{\bf 3} {\bf 3}}=\left( \begin{array}{ccc}
- {\cal A} K(v^2-B^2)+\tilde{\lambda}_7 \Lambda &-2i {\cal A} K v B &
0 \\
-2i {\cal A} K v B & - {\cal A} K (v^2-B^2) +\tilde{\lambda}_6 \Lambda &
0 \\
0 & 0 & -\frac{3}{2} \tilde{\lambda}_9 s
\end{array} \right)
\end{equation}
The determinant is $-3 \tilde{\lambda}_9 s \left({\cal A} K \left( {\cal A} K
(v^2+B^2)^2- (\tilde{\lambda}_6+ \tilde{\lambda}_7)
(v^2-B^2) \Lambda \right) + \tilde{\lambda}_6 \tilde{\lambda}_7 \Lambda^2 
\right)/2$ and is non-vanishing. 
The size of the three masses is expected to be $m_3 \sim \Lambda^2/M$.
For the numerical
solution (I) in Equation \ref{vevsol}, this determinant is $-(0.04)^3$.

The $S$ field contains $({\bf 6},{\bf 1},2/3)+$h.c. and
$({\bf 1},{\bf 3},-1)$+h.c.. These fields acquire Dirac masses $-{\cal A} K
(u^2+A^2)$ and $-{\cal A} K (v^2+B^2)$, respectively. The 
$({\bf \overline{3}},{\bf 1},1/3)+$h.c. and 
$({\bf 1},{\bf 2},-1/2)+$h.c. fields in
the ${\bf 16}+\overline{{\bf 16}}$
acquire Dirac masses 
$-4 \tilde{\lambda}_4 (a^{\prime \prime}+b^{\prime \prime})$
and $-2\tilde{\lambda}_4 (3 a^{\prime \prime}+b^{\prime \prime})$
, respectively.

Finally, there are 8 gauge singlets in this model. 
The quantum modified constraint implies that only 7 of these are independent.
The 
quantum modified constraint can be used to solve for one of the 
gauge singlets. Of the remaining 7, one of these is the Nambu-Goldstone 
boson multiplet of the $SO(10) \rightarrow SU(5)$ symmetry breaking. 
The mass matrix for the remaining 6 gauge singlets is rather 
cumbersome and is not presented here. For the numerical solution 
(I) presented at the start of this Appendix, I have checked that the 
determinant of this matrix is $-6 \times 10^{-7}$ $(\hbox{in units of }
\Lambda=1.)$; the typical 
mass of each singlet is then $\sim 0.09 \Lambda$.

\end{document}